%% file: main.tex
\PassOptionsToPackage{dvipsnames}{xcolor}
\documentclass[acmsmall,screen,nonacm]{acmart}

\usepackage{algpseudocode}
\usepackage{listings}
\usepackage{xcolor}
\usepackage{xspace}
\usepackage[capitalise]{cleveref}
\usepackage{enumitem}
\usepackage{microtype}
\usepackage{fancybox}
\usepackage{graphicx}
\usepackage{subcaption}
\usepackage[frozencache,cachedir=.]{minted}
\usepackage[dvipsnames]{xcolor}
\usepackage{tcolorbox}
\usepackage{listings}
\usepackage{color}

\definecolor{dkgreen}{rgb}{0,0.6,0}
\definecolor{gray}{rgb}{0.5,0.5,0.5}
\definecolor{mauve}{rgb}{0.58,0,0.82}

\definecolor{lightcyan}{rgb}{0.88, 1.0, 1.0}
\setminted[python3]{fontsize=\footnotesize,xleftmargin=8pt,numbersep=5pt,linenos,texcomments,breaklines}

\newcommand{\tool}{\textsc{MPDetector}\xspace}

\newcommand{\Code}[1]{$\mathtt{#1}$}

\renewcommand{\paragraph}[1]{\vskip 0.05in \noindent\textbf{#1.}}

\setcopyright{acmlicensed}
\copyrightyear{2018}
\acmYear{2018}
\acmDOI{XXXXXXX.XXXXXXX}

\acmConference[ISSTA '25]{The ACM SIGSOFT International Symposium on Software Testing and Analysis}{June 25--28,
  2025}{Trondheim, Norway}
\acmISBN{978-1-4503-XXXX-X/18/06}

\begin{document}

% \title[PyDocCon]{PyDocCon: Detecting Inconsistencies Among Multiple Parameters in Python Code and Documentation\cz{Should we position the work as ``API Doc error detection'' rather than ``inconsistency'', as the errors are mostly in doc?}}

% correct one
% \title[MPChecker]{Identifying Multi-Parameter Constraint Errors in Python Data Science Library API Documentations}

\title[MPDetector]{Detecting Multi-Parameter Constraint Inconsistencies in Python Data Science Libraries}

\author{Xiufeng Xu}
\affiliation{%
  \institution{Nanyang Technological University}
  \city{Singapore}
  \country{Singapore}}
\email{xiufeng001@e.ntu.edu.sg}

\author{Fuman Xie}
\affiliation{%
  \institution{The University of Queensland}
  \city{Brisbane}
  \state{Queensland}
  \country{Australia}}
\email{fuman.xie@uq.edu.au}

\author{Chenguang Zhu}
\affiliation{%
  \institution{The University of Texas at Austin}
  \city{Austin}
  \country{USA}}
\email{cgzhu@utexas.edu}

\author{Guangdong Bai}
\affiliation{%
  \institution{The University of Queensland}
  \city{Brisbane}
  \state{Queensland}
  \country{Australia}}
\email{g.bai@uq.edu.au}

\author{Sarfraz Khurshid}
\affiliation{%
  \institution{The University of Texas at Austin}
  \city{Austin}
  \country{USA}}
\email{khurshid@ece.utexas.edu}

\author{Yi Li}
\affiliation{%
  \institution{Nanyang Technological University}
  \city{Singapore}
  \country{Singapore}}
\email{yi\_li@ntu.edu.sg}



%\renewcommand{\shortauthors}{Xiufeng Xu, Yi Li et al.}

\begin{abstract}
  % \cz{I don't like this abstract. I think our storyline should be 1. Data science and machine learning libraries play a crucial role in AI/ML development. 2. These libraries often have APIs with multiple interdependent parameters. 3. Managing these dependencies is challenging and can lead to issues with API compatibility and reliability. 4. To address this challenge, we propose our technique. 5. Our results demonstrate the effectiveness of this approach. We should reorganize both the abstract and intro with this storyline.
  % }

Modern AI- and Data-intensive software systems rely heavily on data science and machine learning libraries that provide essential algorithmic implementations and computational frameworks. These libraries expose complex APIs whose correct usage has to follow constraints among multiple interdependent parameters.
Developers using these APIs are expected to learn about the constraints through the provided documentations and any discrepancy may lead to unexpected behaviors.
However, maintaining correct and consistent multi-parameter constraints in API documentations remains a significant challenge for API compatibility and reliability.
To address this challenge, we propose \tool, for detecting inconsistencies between code and documentation, specifically focusing on multi-parameter constraints.
\tool identifies these constraints at the code level by exploring execution paths through symbolic
execution and further extracts corresponding constraints from documentation using large language
models (LLMs).
We propose a customized fuzzy constraint logic to reconcile the unpredictability of LLM outputs
and detects logical inconsistencies between the code and documentation constraints.
We collected and constructed two datasets from four popular data science libraries and evaluated \tool on them. The results demonstrate that \tool can effectively detect inconsistency issues with the precision of 92.8\%.
% \tool outperforms baseline techniques XXX by XXX.
We further reported 14 detected inconsistency issues to the library developers, who have confirmed 11
issues at the time of writing.

\end{abstract}





\maketitle

\input{content/01.intro.tex}
\input{content/03.background.tex}
\input{content/04.methodology.tex}

\input{content/05.evaluation.tex}
\input{content/06.validity.tex}
\input{content/07.related.tex}

\input{content/08.conclusion.tex}

% \begin{acks}
%   xxx
% \end{acks}

\bibliographystyle{ACM-Reference-Format}
\bibliography{main}
\end{document}

%% file: content/01.intro.tex
\section{Introduction}\label{sec:intro}

% \cz{ I think our storyline should be 1. Data science and machine learning libraries play a crucial role in AI/ML development. 2. These libraries often have APIs with multiple interdependent parameters. 3. Managing these dependencies is challenging and can lead to issues with API compatibility and reliability. 4. To address this challenge, we propose our technique. 5. Our results demonstrate the effectiveness of this approach. We should reorganize both the abstract and intro with this storyline.}

Machine learning (ML) and artificial intelligence (AI) have consistently garnered widespread
attention, achieving remarkable breakthroughs in diverse domains including natural language
processing, recommendation systems, autonomous vehicles, and robotics.
Behind the rapid advancement of these transformative technologies, data science and machine
learning libraries play a crucial role in AI and ML development.
By providing extensive APIs for complex mathematical operations and algorithmic implementations,
these libraries enable researchers and practitioners to focus on solving domain-specific problems
rather than reimplementing fundamental algorithms.

A well-designed API documentation not only provides detailed descriptions of interfaces,
including the purpose and range of parameters or attributes, returns, and exceptions thrown, but may also specify logical constraints or dependencies among multiple parameters.
For data science and machine learning libraries, multi-parameter constraints are commonly mentioned
in their API documentations and users of these libraries are expected to follow them closely when
using the APIs.
%Ideally, these constraints are expected to remain unbroken during actual code execution.
However, frequent version updates may lead to documentations out of sync with the corresponding
code, known as the \underline{\textbf{C}}ode-\underline{\textbf{D}}ocumentation
\underline{\textbf{I}}nconsistency (CDI) issue~\cite{aghajani2019software, rai2022review}.
Such CDI issues are particularly pronounced in data science libraries.
On one hand, the underlying mathematical models of DS/ML libraries inherently come with various
constraints, such as \emph{``a model X can only be chosen when a parameter Y is provided''}, and
incorrect parameter configurations not satisfying their constraints may lead to unexpected outcomes.
On the other hand, the number of parameters/attributes of these libraries can be significantly more
than a typical library API, sometimes over a few dozen.
Therefore, it is unrealistic to track all parameter constraints manually.

Detecting errors in multi-parameter constraints from Python API documentations is challenging for
several reasons.
(1) The quality of API documentations varies and lacks standardized writing guidelines.
Some API documentations use ambiguous languages, contain typos, and may not follow a consistent
styling guide.
This makes simple rule-based pattern matching approaches ineffective.
(2) Existing approaches~\cite{ratol2017detecting, liu2020automating} for detecting documentation errors focus on single parameter
only: e.g., checking whether the information provided on parameter ranges, nullness, and identifier
names is correct.
It is more challenging to extract multi-parameter constraints precisely from free-style descriptions
written in natural languages.
(3) For the same reason, a semantic-aware code analysis approach is essential, as logical relations among
multiple parameters cannot be easily identified through purely syntactic analysis. The challenge is further compounded by Python's dynamic nature, where variable types, attributes, and behaviors can change at runtime.

To detect multi-parameter constraint inconsistencies from data science library documentations, we
propose an automated tool \tool.
\tool identifies inconsistencies between API documentations and the corresponding library code by
combining symbolic execution-based program analysis techniques with constraint extraction methods
powered by large language models (LLMs).
We first extract multi-parameter constraints from documentations (a.k.a. \emph{doc-constraints}),
leveraging the powerful natural language understanding capability of LLMs.
We incorporate a few optimizations, such as Chain of Thought (CoT)~\cite{wei2022chain} and few-shots
learning to improve the accuracy of constraint extraction.
Then we use dynamic symbolic execution to collect all path constraints from the corresponding
Python source code.
The symbolic path constraints (a.k.a. \emph{code-constraints}) capture the real constraints that
the parameters have to follow according to the library code, which are then used to evaluate the
correctness of the doc-constraints.
%In the documentation constraint extraction, we construct prompts applying the  approaches to
%extract constraints, incorporating important information such as parameter
%lists to decrease incorrect outputs and increase the precision of extraction.
%We also provides some fuzzy words to help LLMs deal with some difficult situations to reduce
%omission.

%All these extracted constraints will be sent to a fuzzy constraint converter to be transformed
%into
%fuzzy expressions.
%It helps in handling fuzzy words.
Then, in order to mitigate minor discrepancies that may arise from the doc-constraints extracted by
LLMs, we design and implement a \emph{Fuzzy Constraint Logic} (FCL) framework to estimate how
logically consistent a doc-constraint is with a set of given code-constraints.
Intuitivelly, with the absence of LLM unpredicatability, a doc-constraint must be evaluated to true
under the assumption of code-constraints.
Through fuzzy constraint satisfaction, we can accommodate many \emph{nearly-correct} constraints
produced by LLMs and thus improve the accuracy of the overall approach.
%Finally, we use an SMT-based reasoner to verify whether the constraints are violated in the actual
%execution of the code and infer inconsistencies.

\paragraph{Contributions}
Our work aims to integrate precise symbolic reasoning with the inherently fuzzy outputs of large language models.
To summarize, we make the following contributions.
\begin{enumerate}
    \item We proposed an automated multi-parameter code-documentation inconsistency detection technique and developed an end-to-end command-line tool called \tool. Existing techniques in the same area are only designed to handle single parameter inconsistencies, without considering inter-parameter constraints.
    \item We introduced a customized fuzzy constraint satisfaction framework to mitigate the
    uncertainties introduced by LLM outputs. We provide a theoretical derivation of the membership
    function based on constraint similarity.
    \item We constructed a documentation constraint dataset containing 72 real-world constraints sourced from widely used data science libraries, and a mutation-based inconsistency dataset with 216 constraints generated through systematic mutations.
    Our dataset and tool implementation are made available online: \url{https://anonymous.4open.science/r/MPDetector-E30D}
    \item We evaluated our tool on four real-world popular data science libraries. We reported 14 inconsistency issues discovered by \tool to the developers, who have confirmed 11 inconsistencies at the time of writing.
\end{enumerate}

%% file: content/03.background.tex
\section{Background}\label{sec:bg}
In this section, we review the essential terminology and background necessary for understanding the
remainder of the paper.
% \section{Examples}\label{sec:eg}
\subsection{Multi-Parameter Constraints}\label{sec:eg}
We use two examples to illustrate inconsistencies between API documentation and corresponding code
caused by multi-parameter interdependence.
Both of them come from real-world open-source Python data science libraries and were successfully
detected by \tool.
In general, there are two types of constraints found in API documentations.
(1) An \emph{explicit constraint} clearly specifies the logical relationship between two or more
interrelated parameters in the documentation.
(2) An \emph{implicit constraint} is unstated or indirectly implied relationship between two or
more interrelated parameters, where the constraint is inferred through contexts or convention
rather than explicitly specified.

\subsubsection{Example 1: Explicit Constraint}\label{sec:eg1}

The first example comes from \texttt{statsmodels}~\cite{github:statsmodels}, which provides a
complement to \texttt{scipy} for statistical computations including descriptive statistical and
estimation and inference for statistical models.
\texttt{Statsmodels} has more than 10K stars on GitHub and is actively maintained.
\Cref{fig:eg1} illustrates an inconsistency caused by an explicit constraint from the class
\textit{AutoReg}.
The relevant portions for the doc- and code-constraints are highlighted.
As mentioned in the documentation of \texttt{deterministic}, the trigger condition for the warning
is that ``trend is not n, \textbf{and} seasonal is not False''.
However, it is apparent that the code-constraint for \texttt{trend} and \texttt{seasonal} (to be
used tegother correctly and avoid any warning) implemented is \textbf{or} instead of \textbf{and}.
One way to fix the documentation is to change ``and'' to ``or''.

\begin{figure*}[t]
    \begin{subfigure}{\linewidth}
        \begin{tcolorbox}[colback=Emerald!10,colframe=cyan!40!black,title=\textbf{Constraint description of \texttt{trend} and \texttt{seasonal} in class \texttt{AutoReg}}]
            {\sffamily \textbf{> deterministic: } DeterministicProcess
            \\
            A deterministic process. If provided, trend and seasonal are ignored. \colorbox{blue!20}{A warning is raised if} \colorbox{blue!20}{trend is not "n" and seasonal is not False.}}
        \end{tcolorbox}
        \label{fig:eg1-doc}
    \end{subfigure}
    \begin{subfigure}{\linewidth}
\begin{tcolorbox}[colback=Salmon!20, colframe=Salmon!90!Black,title=\textbf{Corresponding code snippet in class \texttt{AutoReg}}]
\begin{lstlisting}[escapechar=@]
class AutoReg(tsa_model.TimeSeriesModel):
    def __init__(...):
        if deterministic is not None and @\colorbox{blue!20}{(self.trend != "n" or self.seasonal)}@:
            warnings.warn('When using deterministic, trend must be "n"
                and seasonal must be False.', SpecificationWarning, stacklevel=2)
\end{lstlisting}
\end{tcolorbox}
        \label{fig:eg1-code}
    \end{subfigure}
    \caption{Examples of an explicit constraint from \Code{Statsmodels}.}
    \label{fig:eg1}
    \vspace{-5pt}
\end{figure*}

\subsubsection{Example 2: Implicit Constraint}\label{sec:eg2}

The second example comes from \texttt{scikit-learn}~\cite{github:scikit}, which is a widely-used
(59.9K stars on GitHub) open-source machine learning library in Python, designed to offer simple
and efficient tools for data mining and data analysis.
\Cref{fig:eg2} displays an inconsistency caused by an implicit constraint from the class
\textit{SpectralClustering}.
It is evident that the highlighted part of the documentation only explicitly mentions one parameter
\textit{affinity}, omitting the subject ``gamma''.
More importantly, ``ignore'' is not a specific identifier or value but rather a description of
the program logic---if the parameter \textit{affinity} is set to \textit{nearest\_neighbors}, then
the parameter \textit{gamma} will not be used.
Whereas, above constraint does not faithfully reflect the behavior implemented in code.
According to the code snippet, ``gamma'' is not only ignored within the \textit{nearest\_neighbors}
branch, but also ignored within the \textit{precomputed\_nearest\_neighbors} and
\textit{precomputed} branches.
This indicates that the constraint is inaccurate and demonstrates a form of inconsistency.

For this type of implicit constraint, traditional pattern-based approaches are not able to extract
the doc-constraint correctly, thus fail to detect the inconsistencies.
To solve this issue, we design a constraint language that incorporates fuzzy words, and adopt
few-shot learning to teach LLMs how to generate constraints written in this language (details in
\cref{sec:llm}).
In this case, the doc-constraint should be ``(affinity = "nearest\_neighbors") $\rightarrow$
(ignore(gamma))'', where a special predicate ``ignore(x)'' is used to indicate that a variable
\textit{x} is ignored (see \cref{sec:fuzzword}).

\begin{figure*}[t]
    \begin{subfigure}{\linewidth}
        \begin{tcolorbox}[colback=Emerald!10,colframe=cyan!40!black,title=\textbf{Constraint description of \texttt{gamma} and \texttt{affinity} in class \texttt{SpectralClustering}}]
            {\sffamily \textbf{> gamma} : float, default=10
            \\
            Kernel coefficient for rbf, poly, sigmoid, laplacian and chi2 kernels. \colorbox{blue!20}{\textbf{Ignored for}} \colorbox{blue!20}{\textbf{affinity="nearest\_neighbors".}}}
        \end{tcolorbox}
        \label{fig:eg2-doc}
    \end{subfigure}
    \begin{subfigure}{\linewidth}
\begin{tcolorbox}[colback=Salmon!20, colframe=Salmon!90!Black,title=\textbf{Corresponding code snippet in class \texttt{SpectralClustering}}]
\begin{lstlisting}[escapechar=@]
class SpectralClustering(ClusterMixin, BaseEstimator):
    def fit(self, X, y=None):
        if @\colorbox{blue!20}{self.affinity == "nearest\_neighbors"}@:
            ...
        elif @\colorbox{blue!20}{self.affinity == "precomputed\_nearest\_neighbors"}@:
            ...
        elif @\colorbox{blue!20}{self.affinity == "precomputed"}@:
            ...
        else:
            params = self.kernel_params
            if params is None:
                params = {}
            if not callable(self.affinity):
            @\colorbox{blue!20}{params["gamma"] = self.gamma}@
                params["degree"] = self.degree
                params["coef0"] = self.coef0
\end{lstlisting}
\end{tcolorbox}
        \label{fig:eg2-code}
    \end{subfigure}
    \caption{Examples of implicit constraint from \texttt{Scikit-learn}.}
    \label{fig:eg2}
    \vspace{-5pt}
\end{figure*}

\subsection{Fuzzy Logic and Fuzzy Constraint Satisfaction}
Unlike traditional Boolean logic, fuzzy logic~\cite{kosko1993fuzzy} is a multi-valued logic that
allows for values between 0 and 1 to represent varying degrees of truth, where 0 represents
absolute false, and 1 represents absolute true.
The human brain can process vague statements or claims that involve uncertainties or subjective
judgments, such as ``the weather is hot'', ``that man runs so fast'', or ``she is beautiful''.
Unlike computers, humans possess common sense, allowing them to reason effectively in situations
where things are only partially true.
Fuzzy logic is primarily used to model uncertainty and vagueness, making it highly applicable in
real-world scenarios where precision may be difficult or impossible to achieve.

Traditional constraint satisfaction problem (CSP)~\cite{brailsford1999constraint} requires all constraints to be fully
satisfied.
Constraints are either completely satisfied or unsatisfied, which is why these strict,
non-fuzzy constraints are referred to as ``\emph{crisp constraints}''.
An extension of CSP, known as soft CSP~\cite{meseguer2006soft, schiex1992possibilistic}, introduces a distinction between hard
constraints and soft constraints.
Hard constraints must be absolutely satisfied, while soft constraints are typically assigned a
weight or priority, allowing for lower-weighted constraints to be only partially satisfied or even
unsatisfied under certain conditions during problem-solving.
Another extension is Fuzzy CSP~\cite{ruttkay1994fuzzy}, which differs from soft constraints in that it
incorporates fuzzy logic and allows each constraint to be ``partially satisfied'' to a degree,
quantified by a ``satisfaction degree''.
This satisfaction degree usually ranges from 0 to 1, indicating the extent to which a constraint is
fulfilled.
The goal in fuzzy constraint satisfaction is to find a solution that maximizes satisfaction, rather
than strictly satisfy all constraints.

%% file: content/04.methodology.tex
\section{Methodology}\label{sec:method}

\begin{figure*}[h]
    \centering
    \includegraphics[width=\linewidth]{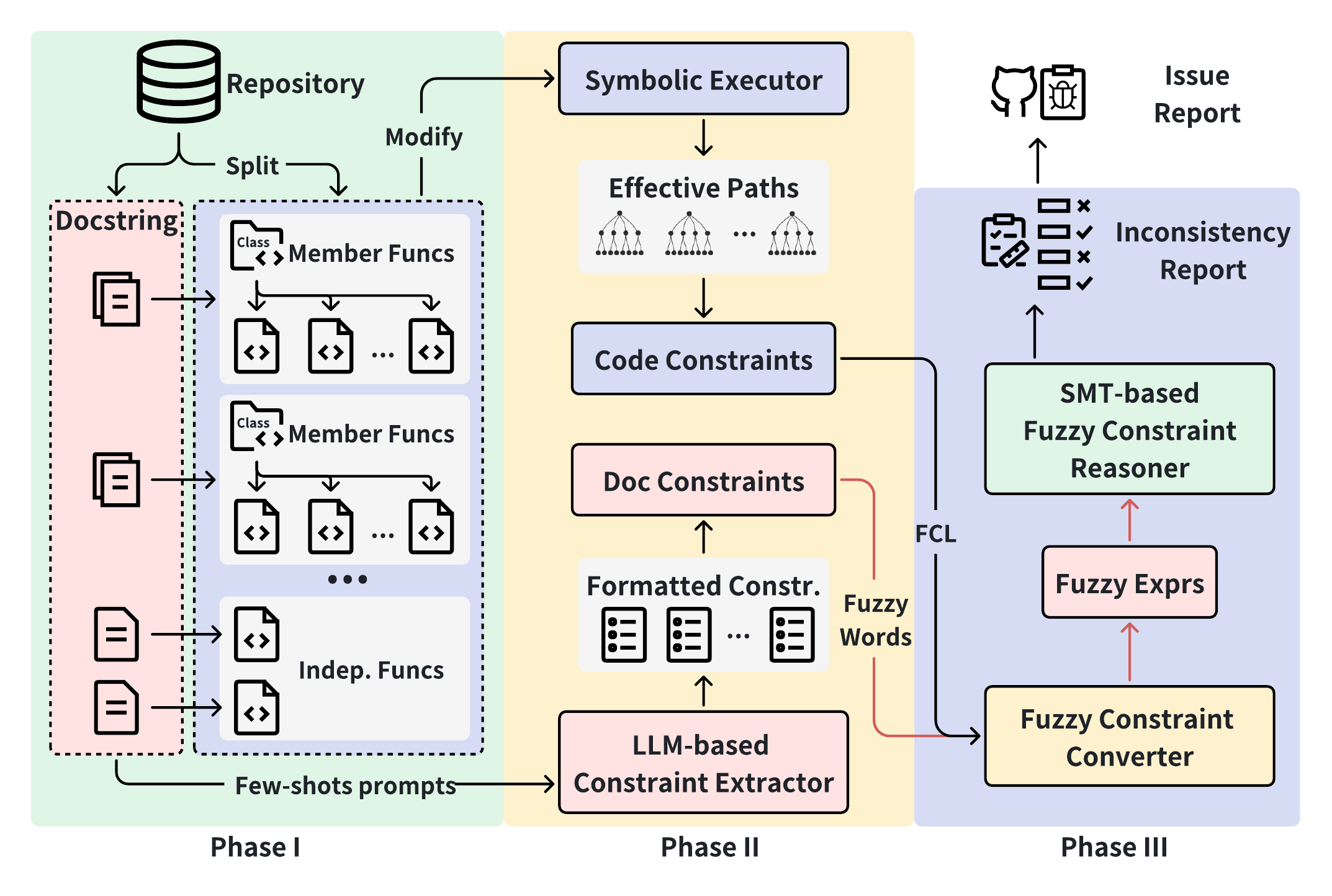}
    \caption{The architectural overview of \tool.}
    \label{fig:arch}
    \vspace{-5pt}
\end{figure*}

In this section, we define the issue of code-documentation inconsistency caused by multi-parameter
constraint and provide a detailed description of our approach. An API documentation error is an inconsistency between the library source code and its API documentation. Multi-parameter constraints refer to conditional dependency relationships that exist among multiple parameters within functions or classes. If a constraint is never violated across all execution paths in the code, it is considered as a benign constraint, or it indicates a potential documentation error. According to literature~\cite{uddin2015api,zhou2017analyzing, zhu2022identifying}, API documentation inconsistency can be categorized into two types: incorrectness and incompleteness. Incorrectness refers to cases where the documentation describes behavior that is not implemented in the code, while incompleteness arises when certain code behaviors are not reflected in the documentation. Typically, incorrectness issues are considered more critical than incompleteness.

In addition, when it comes to constraint extraction, compared to single-parameter constraints, we need to classify the multi-parameter constraint extraction problem into two types, as discussed in Section~\ref{sec:eg}, 1) explicit constraint and 2) implicit constraint.

\tool aims to accurately extract multi-parameter constraints from API documentation and detect both types of inconsistency. As the architecture depicted in Figure~\ref{fig:arch}, we have designed a three-phase workflow comprising the \textbf{1) Data Preprocessing; 2) Constraint Extraction; 3) Inconsistency Detection}. During the preprocessing phase, we separate the code and documentation within the project. We modified each function to facilitate symbolic execution more effectively, and these modifications do not affect the constraint paths of the code. In the constraint extraction and expression generation phase, on the one hand, we leverage large language models to extract constraints in a specific format from the documentation. On the other hand, the symbolic execution tool dynamically analyzes the code and solves the constraint paths. Those constraints are then converted into expressions that can be processed by SMT solver. In fuzzy constraint checking phase, a constraint checker with SMT solver and fuzzy constraint reasoner performs comprehensive reasoning to detect inconsistencies. It is worth noting that we propose and implement a extended fuzzy constraint satisfaction to mitigate the hallucination issues often introduced by large language models, and reduces the risk of false positives and missed detections.

\input{content/041.preprocessing.tex}

\input{content/042.extracting.tex}
\input{content/043.detecting.tex}

%% file: content/041.preprocessing.tex
\subsection{Preprocessing}\label{sec:preprocess}
In this step, we will discuss the details of separating the documentation and corresponding code from the project and the specifics of preprocessing the documentation content.

In modern data science libraries, documentation is typically auto-generated using Sphinx, a tool that can automatically create HTML documentation from Python code. Sphinx supports various docstring styles, with Google style and NumPy style being commonly used. Figures~\ref{fig:docstring} respectively display docstring examples of two different styles from Sphinx official website~\cite{sphnix:google,sphnix:numpy}. Google style docstrings use a clear and concise format with a minimalistic structure. It divides the docstring into sections like \textbf{Args}, \textbf{Attributes}, and etc., with each section using plain indentation. Similarly, Numpy style docstrings organize sections more rigidly. Sections are divided by using \textbf{Parameters}, \textbf{Attributes}, and etc. with horizontal dash lines ``\textbf{- - -}'' under the section header. The number of dash is the same as the number of letters in the section header. Regardless of the style used, the docstring is normally placed at the beginning inside its corresponding class or function.

After downloading the project, our tool first converts every python file from the project into an Abstract Syntax Tree (AST) and isolates the classes and independent functions. This paper focuses on the CDI issue, so in this step, we filter out code without documentation and separately extract the code and documentation from the remaining code. Since Python supports object-oriented programming but current symbolic execution tools have limited support for classes, we have to limit our experimental units to functions. For independent functions, the scope of constraints in the documentation usually applies within the function itself. For classes, however, the constraints cover the entire class, including each member function. Therefore, we create a new directory for every class and independent function, with member function directories placed within their corresponding class directories to maintain structural consistency. If the member function has its own documentation, it will also be retained. 

To help the LLM better focus on the constraints between parameters and reduce the occurrence of erroneous constraints, we retained parameters or attributes and their corresponding descriptions in the form of key-value pairs, based on the two aforementioned docstring styles. We then further applied a rule-based heuristic approach to keep documentations that potentially contain constraints and discard the rest. For instance, if none of the parameters or attributes appear in other descriptions, then there are no multi-parameter constraints in that documentation.

\begin{figure*}[t]
    \begin{subfigure}{0.49\linewidth}
        \begin{tcolorbox}[colback=Emerald!10,colframe=cyan!40!black,title=\textbf{Numpy Style Docstrings}]
\begin{lstlisting}[escapechar=@]
class ExampleNumpyStyle():
  """Exceptions are documented
  @\colorbox{blue!20}{Parameters}@
  ----------
  msg : str
    Human readable string describing the exception.
  code : obj:`int`, optional 
    Numeric error code.
  @\colorbox{blue!20}{Attributes}@
  ----------
  msg : str
    Human readable string describing the exception.
  code : int
    Numeric error code.
  """
  def __init__(self, msg, code):
    self.msg = msg
    self.code = code
\end{lstlisting}
        \end{tcolorbox}
        \label{fig:numpy}
    \end{subfigure}
    \begin{subfigure}{0.49\linewidth}
\begin{tcolorbox}[colback=Salmon!20, colframe=Salmon!90!Black,title=\textbf{Google Style Docstrings}]
\begin{lstlisting}[escapechar=@]
class ExampleGoogleStyle():
  """Exceptions are documented
  Note:
    Do not include the `self` parameter
    in the ``Args`` section.
        
  @\colorbox{blue!20}{Args:}@
    msg (str): Human readable string describing the exception.
    code (:obj:`int`, optional): Error code.

  @\colorbox{blue!20}{Attributes:}@    
    msg (str): Human readable string describing the exception.
    code (int): Exception error code
  """
  def __init__(self, msg, code):
    self.msg = msg
    self.code = code
\end{lstlisting}
\end{tcolorbox}
    \label{fig:google}
    \end{subfigure}
    \caption{Example of two docstring styles}
    \label{fig:docstring}
    \vspace{-5pt}
\end{figure*}

%% file: content/042.extracting.tex
\subsection{Constraint Extraction}\label{sec:extraction}

We now specially explain how to extract path constraints from code and convert them into expressions which are solvable by SMT solver, as well as how to use LLM to extract constraints from documentation and transform them into expressions containing fuzzy words.

\subsubsection{Code Constraint Expression Extraction}

The goal of \tool is to verify whether the constraints between multiple parameters in a documentation align with the logic during actual code execution. This requires our tool to understand and analyze deeper constraint relationships. Therefore, we employ symbolic execution to capture as many path conditions as possible and precisely handle complex paths and constraints.

We modified current advanced dynamic symbolic execution tools~\cite{github:pyexsmt, github:pyexz3, github:pysmt, ball2015deconstructing, bruni2011peer} for path exploration. Unfortunately, Supporting dynamic languages like Python is more challenging compared to symbolic execution tools designed for static languages such as Java and C. Despite Python's rapid evolution, symbolic execution tools specifically designed for Python have developed slowly, struggling to keep pace with the growing new syntax and features. This forces us to make reasonable modifications to the source code extracted directly from repositories. However, these modifications must not alter the path constraints of the original code; they should be equivalent code transformations that do not affect path exploration. We mainly made the following modifications to code:

% We drew inspiration from and modified PyExSMT~\cite{github:pyexsmt}, PyExZ3~\cite{github:pyexz3}, PySMT~\cite{github:pysmt} (all of them adopted dynamic symbolic execution~\cite{ball2015deconstructing, bruni2011peer}) to serve as our symbolic execution engine for path exploration. Unfortunately, these tools remain underdeveloped. Supporting dynamic languages like Python is significantly more challenging compared to symbolic execution tools designed for static languages such as Java and C. To make matters worse, while Python has been evolving rapidly, progress on symbolic execution tools tailored for Python has been slow, unable to keep pace with the growing number of new syntax and features. This forces us to make reasonable modifications to the source code extracted directly from repositories. However, these modifications must not alter the path constraints of the original code; they should be equivalent code transformations that do not affect path exploration. We use LibCST~\cite{github:libcst}, a concrete syntax tree parse and serializer library for Python, to modify source code\cz{this is implementation details, not technique}. We mainly made the following modifications to code:

\begin{enumerate}
    \item Current python symbol execution tool can not solve class directly. Therefore, it is necessary to split the class into functions (i.e. member functions). The corresponding member variables also need to be changed and used as symbolic inputs.
    \item Replace complex structures and operations, such as lists and dictionaries, that are difficult to handle and do not affect the path, as well as external function calls that may cause path explosion, with symbolic inputs.
    \item Replace the handling of exceptions and warnings that do not affect the path with \textit{return}.
    \item Add a fixed format of \textit{return} statement to capture concrete values of potential symbols.
    \item Equivalent code implementation replacement to avoid being unable to find useful path constraints due to poor support for some advanced syntax. For example, replace ternary operator to conventional if-else statement.
\end{enumerate}

% \subsubsection{Path Constraints Extraction}
In the limit, \tool strives to explore all feasible paths in a Python function by following these processes: 1) Running the function with specific input to trace a path through the control flow of the function; 2) Symbolic executing the path to determine how its conditions depend on the function's input parameters; 3) Utilizing Z3 to generate new parameter values that guide the function toward paths that haven't been covered yet.

Although \tool supports a certain level of external function call analysis, in complex real-world code, an external function call often corresponds to extra more function calls, leading to path explosion. Furthermore, documentation constraints are usually handled within the target function, so we still prefer not to introduce external function calls and to focus the analysis within the target function. Additionally, similar to the current concolic symbolic exection tools for Python, \tool does not yet provide strong support for theorem of strings. Thus, during the actual execution process, we replace the string with a unique large number, which does not affect the exploration of condition constraints.

\begin{figure*}[t]
    \begin{subfigure}{0.49\linewidth}
        \begin{tcolorbox}[colback=Salmon!20, colframe=Salmon!90!Black,title=\textbf{Original source code}]
\begin{lstlisting}[escapechar=@]
def fit(self, sample_weight):
  if sample_weight is not None and self.strategy == "uniform":
    raise ValueError("Warning Info")
  if sample_weight is not None:
    sample_weight = _check_sample_weight(sample_weight, X)
\end{lstlisting}
        \end{tcolorbox}
        % \caption{Original source code}
        \label{fig:sourcecode}
    \end{subfigure}
    % \rule{0.5pt}{4cm}
    \begin{subfigure}{0.49\linewidth}
        \includegraphics[width=\linewidth]{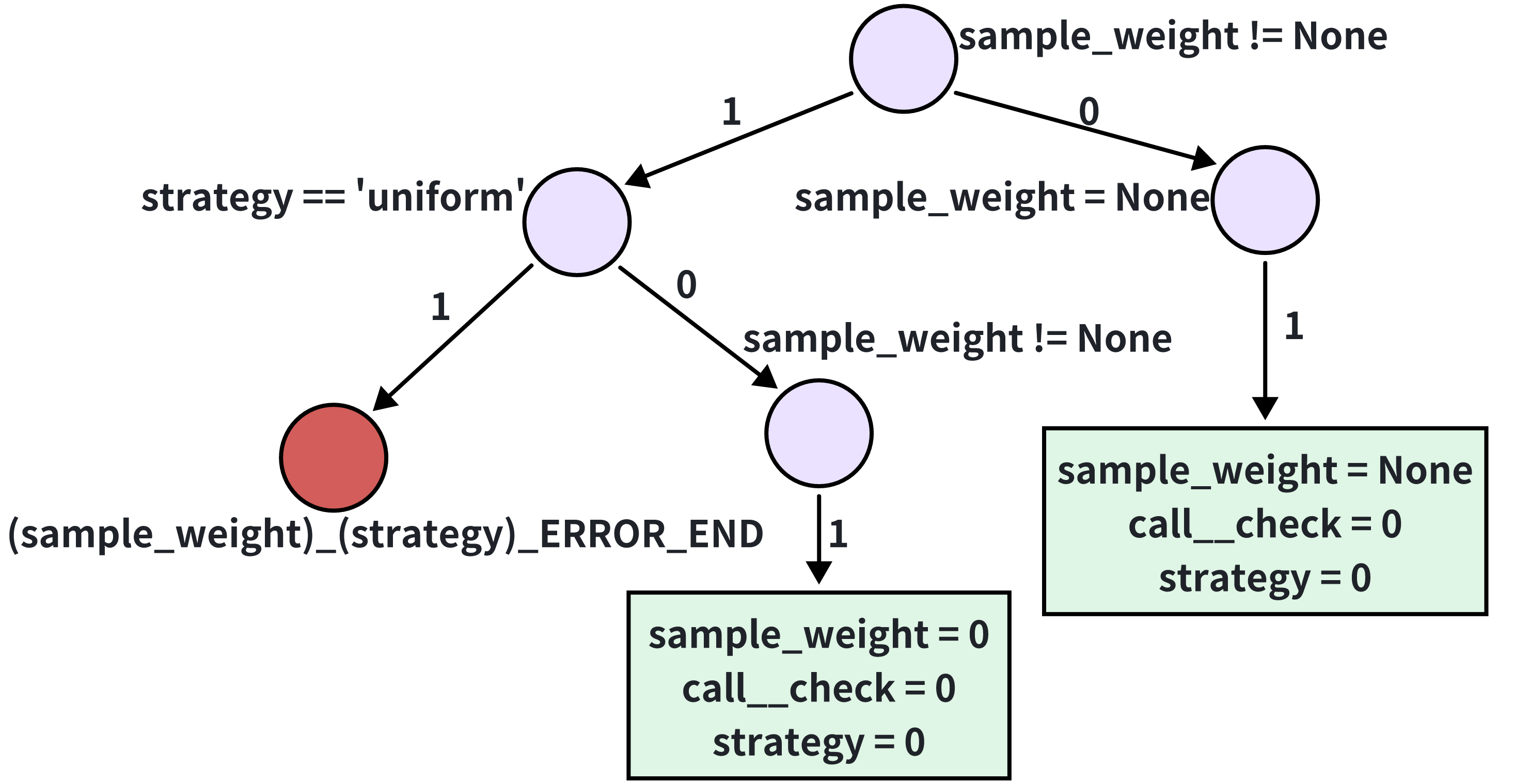}
        \vspace{-8pt}
        % \caption{Path constraint graph}
        \label{fig:pathgraph}
    \end{subfigure}
    \begin{subfigure}{0.98\linewidth}
\begin{tcolorbox}[colback=OliveGreen!10,colframe=Green!70,title=\textbf{Modified source code}]
\begin{lstlisting}[escapechar=@,basicstyle={\footnotesize\ttfamily}]
def fit(sample_weight, strategy, call__check_sample_weight):
  if sample_weight != 'None' and strategy == 'uniform':
    return '(sample_weight)_(strategy)_ERROR_END'
  if sample_weight != 'None':
    sample_weight = call__check_sample_weight
  return (f'(sample_weight = {sample_weight}) ^ (call__check_sample_weight =                         {call__check_sample_weight}) ^ (strategy = {strategy})')
\end{lstlisting}
\end{tcolorbox}
        \label{fig:modifiedcode}
    \end{subfigure}
    \caption{Extracting constraint from code}
    \label{fig:extractpath}
    \vspace{-5pt}
\end{figure*}

We will use an example depicted in Figure~\ref{fig:extractpath} to illustrate the entire extraction phase, containing a simplified original source code and modified code from a popular data science project \texttt{scikit-learn}, and its corresponding path constraints. The \textit{fit} function is a member function within a class, and thus member variables such as ``self.strategy'' also exist within the code. We also modified ``None'' as a string to to make it easier to be captured, since it is represented as a number 0 during symbolic execution. Our tool first modifies the code and replaces exception handling and external function calls with symbolic inputs, marked as ``ERROR\_END'' and ``call\_'', respectively. For those paths whose final states are ``ERROR\_END'', the final results of the conjunction of the documentation constraint and these paths will be negated during reasoning phase.

\subsubsection{Documentation Constraint Expression Extraction}\label{sec:llm}

In this step, we extract constraints from Python documentation by applying LLMs.
Since Python documentation can vary in quality and may contain informal writing~\cite{rani2021comment}, the important task is to understand the parameter information within the documentation. To achieve this, we resort to state-of-the-art LLMs. Given Python documentation as input, the LLM is asked to first extract constraint-related sentences, and then output them in a standard logical expression format. This inlcudes two steps, model selection and prompt design.

% \subsubsection{Model Selection}
\paragraph{Model Selection} We adopt GPT-4, which is pretrained on a diverse corpus and shows
excellent performance in natural language understanding. Based on our preliminary study, GPT-4's
performance stands out compared to Gemini-1.5~\cite{gemini} and LLaMA-3~\cite{llama3} due to its
ability to capture details, and it is also well-acquainted with the context of code
documentation~\cite{dvivedi2024comparative}.

\begin{figure*}[h]
    \centering
    \includegraphics[width=\linewidth]{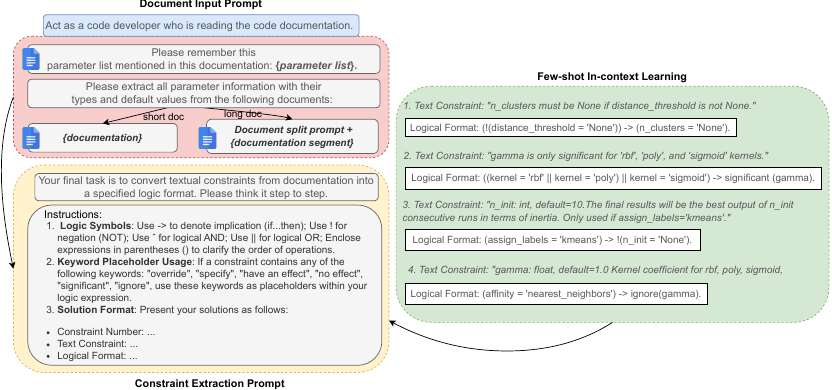}
    \caption{Prompt structure for constraints extraction}
    \label{fig:prompt}
    \vspace{-5pt}
\end{figure*}

% \subsubsection{Prompt Design}
\paragraph{Prompt Design} Because the constraint extraction task is relatively complex and can be
broken down into clear steps, we apply the chain-of-thought approach~\cite{wei2024cot}, which has
been widely proven effective in improving GPT-based model performance. We first divide the prompt
task into two steps, document input and constraint extraction. Figure~\ref{fig:prompt} shows the
structure and some details of the used prompt.
Below, we detail our prompt mechanism for each step.

\paragraph{Document Input Prompt}
We observe that some documentation may be too lengthy to provide to GPT-4 in a single input, considering that GPT-4 have a maximum token length limit of 8,192 tokens~\cite{modeltoken}. We also find that LLMs exhibit lower performance when dealing with long and complex text inputs as noted in previous research~\cite{han2024lm, jin2024llm}. Thus, we decide to segment the lengthy documents into smaller sections. To determine a heuristic chunk size, we randomly select ten lengthy Python documentation files, split them into chunks of varying word lengths, and use these as inputs for GPT-4. We then evaluate the constraint extraction task performance of GPT-4 based on these inputs to determine which chunk size yields better results. Based on our findings, we decide to standardize the chunk size to 1,500 words (around 2,048 tokens~\cite{tokencount}.
We also input the parameter list obtained in Section~\ref{sec:preprocess} into GPT-4 to help model better recognize the information related to parameters. The details of the document input prompt are shown in Prompt 1 in Figure~\ref{fig:prompt}.

\paragraph{Constraint Extraction Prompt}
For the constraint extraction task, our prompt is divided into three parts to guide GPT-4 in recognizing text related to constraints in the original documentation, and then, based on that text, to generate a formatted logical expression of the constraint.

The first part involves defining the logical symbols that can be used in the logical format, including implication, negation NOT, logical AND, logical OR, and also defining parentheses to indicate the precedence of logical expressions.
% The detailed prompt is displayed in Prompt 2(1) in Figure~\ref{fig:prompt}.

The second part arises from our preliminary study, in which we observed that some python documentation uses vague terms such as ``override'', ``specify'', ``have an effect'', ``no effect'', ``significant'', and ``ignore'' when mentioning constraints related to parameters. To preserve as much detail as possible from the documentation, we design prompt to guide GPT-4 so that if text related to parameter constraints contain vague keywords, these keywords should be retained in the final logical expression.
% The detailed prompt is displayed in Prompt 2(2) in Figure~\ref{fig:prompt}.

In the third part, to ensure that the format of the logical expression in GPT-4's output is consistent each time and convenient to process, we apply in-context learning techniques that widely used in previous works~\cite{min2022rethinking, rubin2021learning} to enable GPT-based models to handle tasks specific to a domain.
We include four examples that contain pairs of original constraint-related sentences selected from Python documentation and their corresponding logical expressions constraints.
% The prompt details are shown in Prompt 2(3) in Figure~\ref{fig:prompt}.

%% file: content/043.detecting.tex
\subsection{Inconsistency Detection}\label{sec:detect}

In the second phase, we extracted constraints from both documentation and code. While \emph{code-constraints} are deterministic in nature, \emph{doc-constraints} inherently contain uncertainties stemming from two main sources. First, there are \emph{implicit constraints} arising from vague or incomplete descriptions, which we address by defining fuzzy words to extend them into soft constraints. Second, we encounter uncertainties introduced by generative models' limited reasoning capabilities and unavoidable hallucination issues, for which no validator exists to definitively determine the correctness of generated constraints. To address this challenging scenario, we proposed a customized fuzzy constraint logic to mitigate such vagueness. With the help of fuzzy words and fuzzy constraint logic, our converter can effectively handle both explicit and implicit constraints. The converter ultimately produces fuzzy expressions, which are then processed by a reasoner based on the z3 SMT (Satisfiability Modulo Theories) solver to detect inconsistencies. The principle behind the reasoner is a good constraint means that, for any path constraint $p$ of code, the conjunction of $p$ with the \emph{doc-constraint} $c$ should not result in an unsatisfaction. Naturally, if the end of a path is an exception handling, the result of conjunction of $p$ and $c$ should be addressed oppositely.

% The condition for judgment is defined as follows:

% \begin{equation}
%     \exists p \in P, \neg (p \wedge c)
% \end{equation}

\subsubsection{Fuzzy Words}\label{sec:fuzzword}

% \begin{table}[]
% \caption{Common fuzzy words of two types}
% \begin{tabular}{cc}
% \hline
% \multicolumn{1}{c}{\textbf{Type}}          & \multicolumn{1}{c}{\textbf{Key Words}}                  \\ \hline
% Exsitence     & significant, specify, have an effect, exist \\
% Non-existence & ignore,  no effect, unused, override                      \\ \hline
% \end{tabular}
% \label{tab:fuzzwords}
% \end{table}

The example from Section~\ref{sec:eg2} illustrates a very typical implicit constraint with fuzzy words, where part of constraint are clearly defined while others remain uncertain. We introduce a series of fuzzy words to help LLM extract constraints better. These fuzzy words frequently appear in documentation but don't represent specific values, making it challenging for the LLM to extract them directly. We generally categorize these fuzzy words into two types: \textit{existence} and \textit{non-existence}. In fuzzy words, \textit{non-existence} includes ``ignore'', ``no effect'', ``unused'', ``override'', indicating that a parameter either is unused does or not exist within code segments where other conditions are met. Similarly, \textit{existence} includes ``specify'', ``have an effect'', ``exist'', ``significant'', indicating that the parameter is used or exist when other conditions are met.

We implement some special predicates to evaluate certain relationships. For instance, ``ignore(x)'' are used to check whether a variable \textit{x} is ignored in some conditions or not. \tool will first perform conditional program slicing~\cite{canfora1998conditioned, xu2005brief, weiser1984program, agrawal1990dynamic} on the target code based on explicit conditions in the constraint, extracting the relevant execution paths and code slices under specific conditions. Then, \tool will conduct a data flow analysis on this program slice to generate define-use chain (DU chain)~\cite{kennedy1978use, harrold1994efficient} and attempt to locate the variable in question. If the variable is used, it will return \textit{True}; otherwise, it will return \textit{False}. In real programs, one variable might be assigned to another, so using the DU chain is meaningful in this context.

\subsubsection{Fuzzy Constraint Satisfaction}\label{sec:fuzzlogic}

While we have employed several strategies in Phase II to maximize LLM's understanding of
constraints and restrict randomness in outputs, inaccurate extraction is still unavoidable.
The main reasons are threefold: (1) typos inevitably occur when developers write documentation; (2)
the hallucination issue inherent to black-box generative models; and (3) the intrinsic ambiguity in
natural language.
This implies that correct documentation descriptions can generate incorrect doc-constraints and
vice versa.

With the absence of LLM unpredicatability, detecting CDI issues is a crisp constraint satisfaction
problem (CSP), deciding whether a doc-constraint is aligned with the actual code implementation.
Nevertheless, due to minor errors made by LLMs, such as a single letter being wrongly spelled in a
parameter name, or a comparison operator being reversed, e.g., writing ``<'' instead of ``>'', a
doc-constraint may be deemed as incorrect when it is correct.

To address this, we proposed a customized fuzzy constraint logic that reconcile such
unpredicatability.
In a traditional fuzzy constraint~\cite{kosko1993fuzzy, ruttkay1994fuzzy}, a membership function
assigns a degree of satisfaction (ranging from 0 to 1) to each possible variable value.
It enables partial fulfillment of a condition, with satisfaction measured on a continuous scale.
In our case, a constraint needs to be measured on a new scale, assessing ``how likely'' the
extracted doc-constraint is aligned with the code-constraint.
Therefore, we introduced a unique similarity computation which serves as the membership function.

\Cref{fig:ebnf} shows an EBNF gramma for our multi-parameter constraints.
A multi-parameter constraint is a combination and nesting of binary expressions and Boolean
operators, which can be viewed as a complete binary tree, where leaf nodes are binary expressions
over single parameters and non-leaf nodes are logical operators connecting them.
%Meanwhile, both leaf node and non-leaf node have two variations, corresponding to ``NOT'', and
%``AND/OR'' respectively.
Without loss of generality, we only keep negation, conjunction, and disjunction in the constraints;
logical relations such as implications can be simplified accordingly.
The fuzziness of a constraint is defined with respect to a set of \emph{environment expressions},
facts that are known to hold (with a truth value of 1).
In other words, the instantiation of a specific tree structure and nodes is a constraint $c$
evaluated against a set of expressions $\{e_1, e_2, \ldots, e_n\}$.
Next, we define the membership function of our fuzzy constraint logic through a few similarity
functions.

\begin{figure}[t]
\small\centering
\begin{align*}
   c \in Constraint &::= e \; |\; \neg c \;|\; c \vee c \;|\; c \wedge c \nonumber\\
   e \in Expression &::= \; p \bowtie v\\
   p \in Parameter &::= \; char,\{char \;|\; digit\} \nonumber\\
   \bowtie \; \in Operator &::= \; \textbf{<} \;|\; \textbf{>} \;|\; \textbf{<=} \;|\;
   \textbf{>=} \;|\; \textbf{=} \;|\; \textbf{!=} \; \nonumber\\
   v \in Value &::= \; string \;|\; number \;|\; bool \nonumber
\end{align*}%
\caption{Extended Backus-Naur form for multi-parameter constraint.}\label{fig:ebnf}
\vspace{-5pt}
\end{figure}

\begin{definition}[Expression Similarity]
The similarity between two expressions $e_1$ and $e_2$ is defined as,
\begin{equation}\label{eq:simexpr}
  \sigma(e_1, e_2) = \alpha * (1 - \frac{LD(p_1, p_2)}{\max(|p_1|, |p_2|)}) + \beta *
  (\frac{\delta_{\bowtie_1} \cdot \delta_{\bowtie_2}}{\|\delta_{\bowtie_1}\|
  \|\delta_{\bowtie_2}\|}) + \alpha * (1 - \frac{LD(v_1, v_2)}{\max(|v_1|, |v_2|)}),
\end{equation}
where $\alpha$ and $\beta$ denote relative weights, $p$, $\bowtie$, and $v$
are parameter, operator, and value, respectively, $|p|$ and $|v|$ denotes the length of $p$ and $v$, $\|\delta_{\bowtie}\|$ denotes the magnitudes (or Euclidean norms) of the vector $\delta_{\bowtie}$.
\end{definition}

The similarity between two expressions are considered separately for the parameters, operators, and
values appeared in the experssions.
Both $p$ and $v$ can be treated as texts, therefore, Levenshtien Distance (a.k.a.
edit distance) is used to represent their similarity.
The normalized Levenshtien Distance (NLD) is given in Equation~\ref{eq:ld}, where $s$ denotes strings ($p$ or $v$), $|s|$ denotes the length of it.

\begin{equation}\label{eq:ld}
    \eta(s_1, s_2) = NLD =  1 - \frac{LD(s_1, s_2)}{\max(|s_1|, |s_2|)}
\end{equation}

\begin{equation}\label{eq:embedding}
    \delta_{\bowtie} = (C, E, G, L, N), C, E, G, L, N \in \{0, 1\}
\end{equation}

\begin{equation}\label{eq:cossim}
    cos\theta(\bowtie_1, \bowtie_2) = \frac{\delta_{\bowtie_1} \cdot \delta_{\bowtie_2}}{\|\delta_{\bowtie_1}\| \|\delta_{\bowtie_2}\|}
\end{equation}

The differences between operators
As shown in the Equation~\ref{eq:embedding}, we embeds the operators according to five
characteristics, which are \underline{\textbf{C}}omparison, \underline{\textbf{E}}quation,
\underline{\textbf{G}}reater than, \underline{\textbf{L}}ess than, \underline{\textbf{N}}egativity.
In this way, we can calculate the similarity of two operators by simply calculating the cosine
similarity between two vectors. The result is highly intuitive. For example, with the $\delta_{<}
= (1, 0, 0, 1, 0)$, $\delta_{>}  = (1, 0, 1, 0, 0)$, and $\delta_{<=}  = (1, 1, 0, 1, 0)$, the
similarity between ``$<$'' and ``$>$'' is 0.5, while the similarity between ``$<$'' and ``$<=$'' is
0.82.

We also set $\beta$ as the weight to adjust the sensitivity to changes in operators, and the weights for parameters and values should be the same, so $\alpha = \frac{1 - \beta}{2}$. Therefore, the similarity $\sigma$ of two single parameter expressions (i.e. atomic constraints) can be calculated by Equation~\ref{eq:simexpr}.

\begin{definition}[Constraint Similarity]
The similarity between two constraints $c$ and $\varphi$ can be calculated by a combination of following formulas,
\begin{equation}\label{eq:js}
    \rho(c,\varphi) \Longrightarrow
    \begin{cases}
        \mathop{\arg\max}\limits_{e_i \in \varphi} \sigma(e, e_i), \varphi = \{e_i|i=1,2,\ldots,n\}, & \text{if } c = e \\
        1 - \sigma(c', \varphi), & \text{if $c = \neg c'$} \\
        \min\{\sigma(c_1, \varphi), \sigma(c_2, \varphi)\}, & \text{if } c = c_1 \wedge c_2 \\
        \max\{\sigma(c_1, \varphi), \sigma(c_2, \varphi)\}, & \text{if } c = c_1 \vee c_2 \\
    \end{cases}
\end{equation}
\end{definition}
Consider an atomic constraint with a single expression; its similarity to a constraint $\varphi$ associated with a set of expressions can be represented by the maximum expression similarity among all expressions within $c$. Based on \textit{conjunctive combination principle}~\cite{zadeh1965fuzzy}, when combining two constraints using a conjunction, their degree of joint similarity $\rho$ should be represented by the minimum similarity between them. Similarly, based on \textit{disjunctive combination principle}~\cite{zadeh1965fuzzy}, when they are combined with a disjunction, the maximum similarity should be used. For negation, the complementary similarity will be used.

% \textbf{Definition 2: Constraint Similarity.}
% Based on \textit{conjunctive combination principle}~\cite{zadeh1965fuzzy}, when
% combining two constraints using a conjunction, their degree of joint similarity $\rho$ should be
% represented by the minimum similarity between them.
% Similarly, based on \textit{disjunctive combination principle}~\cite{zadeh1965fuzzy}, when they are combined
% with a disjunction, the maximum similarity should be used.
% For negation, the complementary similarity will be used.
% Formulas are shown in Equation~\ref{eq:js}
% \yi{rewrite using the definition environment as in Def.1}

\begin{definition}[Membership Function]
The membership function that represents the degree to which a constraint $c$ belongs to a fuzzy set is defined as,
\begin{equation}\label{eq:mem}
    % \mu(c) = \frac{1}{n}\sum_{i=1}^{n} \max\{\rho(c, \varphi_i)P(c_i \wedge \varphi_i), P(c \wedge \varphi_i)\}
    \mu(c) = \frac{1}{n}\sum_{i=1}^{n} \rho(c, \varphi_i)P(c_i \wedge \varphi_i)
\end{equation}
where $\mu(c) \in [0,1]$, $\varphi_i$ denotes one of the path constraint in a set $\Phi$ containing all path constraints, $\Phi = \{\varphi_1, \varphi_2, \ldots, \varphi_n\}$; $\rho(c, \varphi_i)$ denotes the degree of joint satisfaction of $c$ and one of path constraint $\varphi_i$; $c_i$ denotes a variation of $c$ where all expressions in $c$ are replaced with the closet matching expressions in $\varphi_i$; the predicate $P(c, \varphi)$ represents whether a given constraint satisfies one of path constraint.
\end{definition}

The value of $P(c, \varphi)$ is either ``True'' or ``False'', corresponding to 1 or 0, respectively. Hence, ``True'' and ``False'' are complementary, for example, ``0.7False = 0.3True''.

% \textbf{Definition 3: Membership function.} Equation~\ref{eq:mem} demonstrates the membership function $\mu(c)$ that represents the degree to which an constraint belongs to a fuzzy set, where $\mu(c) \in [0,1]$. Given a set $\Phi$ containing all path constraints, $\Phi = \{\varphi_1, \varphi_2, \ldots, \varphi_n\}$. Based on two definitions above, the degree of joint similarity $\rho(c, \varphi_i)$ between a documentation constraint $c$ and one of path constraint $\varphi_i$ will be calculated. Additionally, a predicate $P(c, \varphi)$ is adopted to determine whether a given constraint satisfies one of path constraint. The result is either ``True'' or ``False'', corresponding to 1 or 0, respectively. We use constraint similarity when a constraint does not satisfy a specific path constraint.

% \begin{equation}\label{eq:mem}
%     \mu(c) = \frac{1}{n}\sum_{i=1}^{n} \max\{\rho(c, \varphi_i), P(c \wedge \varphi_i)\}
% \end{equation}
% \yi{rewrite using the definition environment as in Def.1}

%% file: content/05.evaluation.tex
\section{Evaluation}\label{sec:eval}
This section describes our evaluation of \tool. We first present our research questions, then detail the experiment setup and evaluation subjects. Finally, we analyze our experimental results and answer each research question. Our evaluation was guided by the following research questions:

\begin{enumerate}
  \item \textbf{RQ1:} How accurate is \tool in extracting constraints from API documentation?
  \item \textbf{RQ2:} How effective is \tool in detecting errors related to multi-parameter constraints in API documentation?
  % \item \textbf{RQ3:} How much impact does each component have on the final performance of \tool?
  \item \textbf{RQ3:} How effective can \tool detect unknown inconsistency issues?
\end{enumerate}

\subsection{Experiment Setup}
\subsubsection{Dataset.}
There is currently no well-established dataset specifically focusing on multi-parameter API
documentation errors. To better evaluate the effectiveness of our tool, we constructed two datasets: a constraint dataset and an inconsistency dataset.

\paragraph{Constraint Dataset}
We curated a constraint dataset containing 72 constraints from 4 popular open-source data science
libraries, including \texttt{scikit-learn}, \texttt{scipy}, \texttt{numpy}, and \texttt{pandas}.
We collected constraints from commits on GitHub to track developers' modifications on documentation
related to multi-parameter constraints.
Frequent modifications are more likely to result in inconsistencies.
We developed an automated script to collect all documentation-related commits and
identify changes within parameter (including attribute) descriptions according to two docstring
styles (see details in Figure~\ref{fig:docstring}). By mapping parameter names to their
descriptions, we then cross-check if any parameter names appear within others' descriptions to
filter out potential constraint-related documentation. This heuristic approach excludes most
irrelevant commits, leaving a subset for manual review to confirm constraints. For each constraint,
we log its source information (repository name, SHA, file path, and etc.) and retain the code file,
enabling swift extraction of documentation and code for reproducibility. Additionally, some
constraints are not checked within a called function rather than the current function, leading to
mismatch between documentation and code. In our dataset, we record these mismatches and place the
constraint descriptions within the function where they are actually checked.

\paragraph{Inconsistency Dataset}
Based on constraint dataset, we constructed an inconsistency dataset comprising 126 multi-parameter
constraints that lead to code-documentation inconsistencies. We analyzed around 20 resolved GitHub
issues related to multi-parameter constraints and identified eight common patterns that may cause
CDI: (1) Parameter name change; (2) Value Change; (3) Logic Change; (4) Remove Parameter; (5) Add
Constraints; (6) Remove Constraints; (7) Missing Documentation; (8) Modify Description. To better
evaluate the capabilities of our tool, we applied these eight patterns to mutate the dataset based
on the \emph{Constraint Dataset}. For each constraint, we applied two types of modifications, resulting in an
inconsistency dataset containing 216 constraints. We feed the mutated constraints and the original
constraints into an SMT solver to verify if the mutations violate the original constraints. We also
manually inspected each them to ensure the constraint is inconsistent. It is important to note that
modifying a correct constraint does not necessarily turn it into an incorrect one. As a result, we
obtained an \emph{Inconsistency Dataset} containing 126 inconsistent constraints and 90 consistent
constraints. In a sense, our dataset can be considered as potential inconsistencies that may
realistically occur during development.

\begin{table}[]
\caption{Data science libraries used in the experiments}
\begin{tabular}{lrrrrrrr}
\hline
\multicolumn{1}{c}{Project} & \multicolumn{1}{c}{class} & \multicolumn{1}{c}{class w/ doc} & \multicolumn{1}{c}{func} & \multicolumn{1}{c}{func w/ doc} & \multicolumn{1}{c}{KLOC} & \multicolumn{1}{c}{avg. params} & \multicolumn{1}{c}{\#Stars} \\ \hline
scikit-learn                & 878                       & 306                              & 9,332                    & 1535                            & 400.1                    & 1.42                            & 59,954                      \\
pandas                      & 2,211                     & 102                              & 28,880                   & 1432                            & 620.6                    & 1.14                            & 43,666                      \\
scipy                       & 2,570                     & 142                              & 22,059                   & 1705                            & 517.9                    & 1.30                            & 13,053                      \\
numpy                       & 2,000                     & 51                               & 12,618                   & 902                             & 276.3                    & 0.83                            & 27,935                      \\ \hline
keras                       & 1,370                     & 254                              & 9,877                    & 724                             & 218.5                    & 1.42                            & 61,969                      \\
dask                        & 284                       & 26                               & 6,914                    & 368                             & 157.5                    & 1.33                            & 12,552                      \\
statsmodels                 & 2,184                     & 273                              & 11,590                   & 1,894                           & 424.6                    & 1.32                            & 10,100                      \\ \hline
\end{tabular}
\label{tab:subjects}
\end{table}

\subsubsection{Subjects}
Table~\ref{tab:subjects} lists 7 popular libraries that \tool evaluated on, first four for dataset construction and last three for assessing \tool's ability to detect unknown issues. The selected libraries are of high quality and widely used, with tens of thousands of stars on GitHub. These libraries are substantial third-party libraries, averaging 1,642 classes, 14,467 functions, with an average of 373.6 thousands of lines of code, and each function containing an average of 1.3 parameters. Our tool extracts documentation constraints and path constraints from these libraries and uses a fuzzy constraint reasoner to detect inconsistencies.

% add and experiment of fuzzy constraint satisfaction
% 我们从收集被报告并且已经被开发团队处理的github issue中 确定了以下常见问题。为了验证fuzzy constriant的效果，我们还在mutation的过程中额外增加了参数，取值，操作符的改变导致的不一致的约束的数量。

\tool was implemented in Python. All the experiment were performed on a Intel(R) Xeon(R) Silver 4214 CPU @ 2.20GHz machine with 252GB of RAM, running Ubuntu 18.04, with Python 3.8.19 and Z3 4.13.0.
When evaluating the constraint extraction performance (RQ1), we access the GPT-4 model through OpenAI's API.
% \footnote{We access it from https://openai.com/index/openai-api/.}.
For result validation, given the absence of established benchmarks in this domain, two volunteer researchers independently reviewed the constraint extraction results from GPT-4, manually assessing each constraint's consistency with the original Python documentation. Any discrepancies were resolved through consensus discussion.

% For evaluating the LLM results, since there is a lack of previous benchmarks, we involved two volunteers from our research team to review the constraint extraction results of GPT-4. For each constraint, the volunteers were asked to manually confirm whether the extracted constraint was consistent with the corresponding text in the original Python documentation. In case of any disagreement, the two volunteers discussed the issue until the conflict was resolved.

% We evaluate the accuracy of constraint extraction by calculating \emph{the percentage of constraints correctly extracted / the total number of constraints} in the benchmark.

% \textcolor[]{red}{i think there also need a paragraph to describe some setup details of rq2?}

\subsection{Results}
\subsubsection{Accuracy of LLM in extracting constraints from API documentation.}
We display the result of RQ1 in Table~\ref{tab:rq1}. Our experiment shows that our tool correctly identified and extracted 66 out of 72 constraints contained in the Python documentation collected in our benchmark, achieving an accuracy of 91.7\%, which demonstrates that our tool can successfully extract most of the constraints accurately, with few errors or omissions.
Next, we look into the remaining failed cases and investigate the reasons for the inaccuracy. We found that out of the 6 incorrect cases, 4 involved missing constraints during the documentation processing. For example, one of the missing constraints is: ``(batch\_size = auto) $\rightarrow$ (batch\_size = min(200, n\_samples))''. Although this case involves a constraint between multiple parameters, the format is tricky because one of the parameters is within a function, which may have misled GPT-4 and caused it to miss this constraint during extraction. In the last 2 cases, the constraint information was identified but converted into incorrect logic expressions due to the complex logic or sentence structure.

We further explore \tool's abilities by conducting an ablation study. The results show that \tool without few-shot learning achieves an accuracy of 62.5\%. Most failures occur in the incorrect extraction of constraints. This indicates that including few-shot learning is important for \tool to generate accurate constraints.
Next, \tool without applying chain-of-thought techniques results in an accuracy of 79.2\%, with the number of missed constraints accounting for more than half of total non-equivalent cases. This suggests that GPT tends to miss more details in documentation when chain-of-thought is removed.
After including chain-of-thought and few-shot learning, \tool's performance shows a clear improvement.

\begin{table}[]
\caption{Results of \tool on constraint extraction}
\begin{tabular}{lcccc}
\hline
\multicolumn{1}{c}{}                    & \textbf{Equivalent}                                                & \multicolumn{2}{c}{\textbf{Non-Equivalent}}                                                                                                 & \textbf{}         \\ \hline
\multicolumn{1}{c}{}                    & \begin{tabular}[c]{@{}c@{}}\# of correct \\ extraction\end{tabular} & \begin{tabular}[c]{@{}c@{}}\# of incorrect \\ extraction\end{tabular} & \begin{tabular}[c]{@{}c@{}}\# of missing \\ constraints\end{tabular} & \textbf{Accuracy} \\ \hline
\textbf{\tool w/o few-shot learning} & 45                                                                 & 20                                                                   & 7                                                                    & 62.5\%              \\ \hline
\textbf{\tool w/o chain-of-thought}  & 57                                                                 & 7                                                                    & 8                                                                    & 79.2\%              \\ \hline
\textbf{\tool}                       & 66                                                                 & 2                                                                    & 4                                                                    & 91.7\%           \\ \hline
\end{tabular}
\label{tab:rq1}
\end{table}

\smallskip
\noindent\shadowbox{%
  \begin{minipage}{0.98\columnwidth}
    \textbf{Answer to RQ1:} \tool correctly extracted 66 constraints out of 72 in total, achieving an accuracy of 91.7\%. This demonstrates that \tool is effective in extracting constraints from Python documentation.
  \end{minipage}}

 % for \todo{xxx} functions

% ===== RQ2 =====
\subsubsection{\tool's effectiveness in detecting multi-parameter API documentation errors}
To measure \tool's effectiveness in detecting multi-parameter API documentation errors, we evaluated our tool on inconsistency dataset. Table~\ref{tab:res} shows the results of \tool in detecting multi-parameter CDI on the inconsistency dataset.

The vanilla \tool does not account for fuzzy words or apply fuzzy constraint satisfaction theory, limiting its ability to handle implicit constraints. Despite these limitations, it achieved an impressive 69\% precision by detecting 89 inconsistencies. With the addition of fuzzy words and fuzzy constraints, \tool's performance significantly improved, successfully identifying 119 inconsistencies with a precision of 92.8\%. This demonstrates that incorporating fuzzy words and fuzzy constraint can expand the range of detectable constraints.

However, we observed two false positives. These two false positives occurred because the
constraints involved were between parameters and method calls, rather than between the parameters.
One example is shown below:
\begin{minted}[highlightlines={5},linenos=false]{python3}
  ((n_nonzero_coefs='None')^(tol='None'))->n_nonzero_coefs_=max(0.1*n_features,1)
\end{minted}
where, the value for ``n\_nonzero\_coefs\_'' is a function ``max''. Modern software increasingly emphasizes code maintainability and reusability, leading to highly complex function calls. To prevent the high risk of path explosion in symbolic execution, we alternate function calls with symbolic inputs during the preprocessing phase, which leads to misclassification of these two inconsistencies.

The remaining 9 unresolved constraints stem from external function dependencies. While incorporating external function code could resolve these constraints, this approach risks infinite recursive dependencies and path explosion. For us, best practices suggest that constraints should be handled within the documented function itself.

\begin{table}[]
  \caption{Results of LLM check and \tool on detecting multi-parameter CDIs}
  \begin{tabular}{lrrrr}
  \hline
    & FP  & TP  & Precision\\ \hline
  LLM Checker        & 117  & 7  & 5.6\% \\
  LLM Checker w/ constraints      & 74  & 52  & 41.3\% \\
  % \tool                & 35  & 33  & 26.2\%\\
    \tool Vanilla   & 2  & 87  & 69.0\%\\
  \tool w/ fuzzy words\&fuzzy constraints & 2 & 117 & 92.8\% \\ \hline
  \end{tabular}
  \label{tab:res}
\end{table}

% 20条constraint 收到不正确文档的影响，28条exist 3ignore 1effect存在fuzz word的判断 重合5
% 15个可以通过简单的启发式方法得到文档

A notable situation arose during one of our issue reporting, even though our issue had been confirmed, we encountered dissatisfaction from one of the developers. He believed we were an automated tool or bot based on AI because of our anonymous status, which diminished his enthusiasm for addressing the issue. With a mass of LLM-based program analysis or inconsistency checkers now available, while they offer insights sometimes, their results often cost more manual verification than traditional tools due to higher uncertainty.

\paragraph{Comparative Study}
Therefore, we conducted a comparative experiment between our tool and the approach of using only LLMs as a constraint checker on the same dataset. To align with our experiment settings, we chose GPT-4, one of the leading models, for comparison and evaluated its performance under two settings: 1) providing the raw documentation and code as input, directly prompting GPT to check for consistency, and 2) providing the extracted constraints along with the code, requesting a consistency check. We also require it to provide justifications for its answers.

As shown in the Table~\ref{tab:res}, when raw documentation and the corresponding code were provided as inputs to the GPT, GPT demonstrated significant limitations in detecting multi-parameter CDIs, finding only 7 inconsistencies with a precision of 5.6\%. When extracted constraints and code were given as inputs, the model demonstrated heightened awareness of the task and had a higher probability of locating the constraint-related code segments. However, it still struggled to determine inconsistency. Out of 126 inconsistent constraints, 52 were identified correctly, yielding a precision of 41.3\%. After a thorough review of GPT's responses, we found that GPT gave correct results in many cases but did not provide a reasonable explanation or even wrong explanation. These results indicate that LLM still has limitations in detecting complicated multi-parameter CDIs.

\smallskip
\noindent\shadowbox{%
  \begin{minipage}{0.98\columnwidth}
    \textbf{Answer to RQ2:}
    Compared to an LLM-only checker with a precision of 41.3\%, \tool successfully detected 117 out of 126 inconsistent constraints, achieving a 92.8\% precision. Notably, the application of fuzzy words and fuzzy constraints improved \tool's precision by 23.8\%.
  \end{minipage}}

\subsubsection{Practical effect of \tool}
Our tool's effectiveness in detecting unknown multi-parameter inconsistencies was validated by manual review and developers feedback. We reported 14 inconsistencies identified by \tool to the library maintenance team, receiving positive engagement and warm responses. Two of them even spark further discussions about potential issues. This not only affirmed our reports' quality but also reflected the enthusiasm of the open-source community.

For example, an issue confirmed by the \texttt{scikit-learn} team originates from the independent function ``lars\_path'', as shown in Figure~\ref{fig:eg3}. Apparently, an inconsistency exists between the documentation and code regarding whether ``Gram'' is None when ``X'' is None. Therefore, we reported the issue~\cite{issue30099} and detailed the documentation sections with inconsistencies alongside its corresponding code snippet. The developer made a bit of archeology, admitted the documentation needs to be updated, and asked do we want make a PR to correct this error. Finally, the documentation description was fixed to ``If X is None, Gram must also be None''.

For most issue reports, we received quick feedback, and 11 inconsistencies were confirmed and improvements were made to documentation or code. Of the remaining three cases, two were reported at the initial phase of our experiments when our understanding of the project architecture was insufficient. The checks for these two constraints are done in other deeper files, but we were unable to verify them accurately at that time. The third inconsistency stemmed from ambiguity in the natural language, which resulted in a different interpretation diverged from the developers' original intent. Furthermore, these reported issues have contributed to four DS/ML repositories (\texttt{scikit-learn}, \texttt{keras},\texttt{statsmodels},\texttt{dask}), which emphasizes the generalization of our tool.

\begin{figure*}[t]
    \begin{subfigure}{\linewidth}
        \begin{tcolorbox}[colback=Emerald!10,colframe=cyan!40!black,title=\textbf{Constraint description of \texttt{X} and \texttt{Gram} in function \texttt{lars\_path}}]
            {\sffamily \textbf{> X } : None or ndarray of shape (n\_samples, n\_features)
            \\
            Input data. Note that \colorbox{blue!20}{\textbf{if X is None then the Gram matrix must be specified}}, i.e, cannot None or False.}
        \end{tcolorbox}
        \label{fig:eg2-doc}
    \end{subfigure}
    \begin{subfigure}{\linewidth}
\begin{tcolorbox}[colback=Salmon!20, colframe=Salmon!90!Black,title=\textbf{Corresponding code snippet in function \texttt{lars\_path}}]
\begin{lstlisting}[escapechar=@]
def lars_path(...):
    @\colorbox{blue!20}{if X is None and Gram is not None:}@
        raise ValueError("X cannot be None if Gram is not None. Use lars_path_gram to avoid passing X and y.")
\end{lstlisting}
\end{tcolorbox}
        \label{fig:eg3-code}
    \end{subfigure}
    \caption{Examples of implicit constraint from \texttt{Scikit-learn}.}
    \label{fig:eg3}
    \vspace{-10pt}
\end{figure*}

\smallskip
\noindent\shadowbox{%
  \begin{minipage}{0.98\columnwidth}
    \textbf{Answer to RQ3:}  We reported 14 multi-parameters inconsistencies detected by \tool to library developers, who have already confirmed 11 inconsistencies by the time of submission (confirmation rate = 78.6\%)~\cite{issue28469,issue28470,issue28473,issue29440,issue29463,issue29464,issue9304,issue29509,issue11336,issue20141, issue30099}. These results demonstrate that \tool can effectively detect unknown API documentation errors. Some of them are even in unseen libraries which highlights its strong generalization capability.
  \end{minipage}}

%% file: content/06.validity.tex
\section{Threats to Validity}\label{sec:valid}

\paragraph{Internal}
There is no established ground truth for multi-parameter code-documentation inconsistencies. To
mitigate this, we manually collected and verified the inconsistencies detected by \tool. Given the
complexity of multi-parameter constraints, two of our authors spent an additional 10 minutes per
inconsistency to verify whether it was a true positive. Furthermore, many of the confirmed true
positive errors have been confirmed by the library developers, which also supports the accuracy of
our manual labeling process.

\paragraph{External}
Our experimental results may not be applicable to other data science
libraries. To address this limitation, we chose high-quality, widely used data science libraries as
representatives, ensuring that both their source code and API documentation were accessible.
Currently, \tool only supports two docstring styles, but in the future, our approach can be
extended to other styles.

%% file: content/07.related.tex
\section{Related Work}\label{sec:related}

\paragraph{API Documentation Analysis}
Numerous empirical studies have revealed the challenges of maintaining high-quality API
documentations~\cite{aghajani2018large,arnaoudova2016linguistic, dagenais2010creating,
aghajani2020software, head2018not, liu2020generating, monperrus2012should, saied2015observational,
shi2011empirical, uddin2015api, zhong2020empirical, kang2021active}. These studies indicate that
documentation errors are prevalent, even in well-established and widely-used libraries.
Additionally, a empirical study from Aghajani et al\cite{aghajani2018large}. shows that linguistic antipatterns in APIs
increase the likelihood of developers introducing errors and raising more questions compared to
using clean APIs. Saied et al.\cite{saied2015observational} conducted an observational study focusing on API usage constraints
and their documentation. Zhong and Su\cite{zhong2013detecting} proposed a method that combines natural language processing
(NLP) with code analysis to identify errors in API documentation, specifically targeting
grammatical mistakes (such as spelling errors) and incorrect code references (i.e., names that do
not exist in the source code). Lee et al.\cite{lee2019automatic} developed a technique to extract change rules from code
revisions and apply them to detect outdated API names in Java documentation, with a particular
focus on names of Java classes, methods, and fields.

Another related field is code Comments\cite{blasi2018replicomment, habib2018class, liu2014automatic, nie2019framework, panthaplackel2021deep, steidl2013quality, zhai2020cpc, wen2019large}. Existing research on code comment analysis predominantly follows two approaches. The traditional method employs program analysis and heuristic rules to detect inconsistencies between comments and code. Technologies like CUP\cite{liu2020automating}, CUP2\cite{liu2021just}, and HebCup\cite{lin2021automated} exemplify this approach, focusing on automatic just-in-time comment updates when corresponding code changes. The alternative approach leverages NLP techniques, particularly LLMs, to retrieve and extract information from software artifacts.

\tool focuses on a distinct problem, specifically on API documentation errors arising from multi-parameter constraints. These issues are more subtle and challenging to detect, particularly within data science libraries built on the dynamic language Python.

\paragraph{LLM-based Program Analysis}
A line of research~\cite{wadhwa2024codequality,jin2023programrepair, yang2024programrepair, xia2024programrepair, nam2024codeunderstand,zhang2024codeinconsistency, zhang2024codeinconsistency} focus on using LLMs on program analysis. Wadhwa et al.~\cite{wadhwa2024codequality} focus on using LLMs to resolve code quality issues on multiple code language. Several recent research~\cite{jin2023programrepair, yang2024programrepair, xia2024programrepair} address applying LLMs on program repairing issue. Nam~\cite{nam2024codeunderstand} apply GPT-3.5-turbo model to explain code and providing usage details.
The existing approaches focus on different purposes compared to \tool.
Zhang~\cite{zhang2024codeinconsistencyfse, zhang2024codeinconsistencyase} uses LLMs to extract constrains from code comments, and applies AST-based Program Analysis to identify inconsistencies.
Rong et al.~\cite{rong2024code} propose C4RLLaMA, a fine-tuned large language model based on the open-source CodeLLaMA, to detect and correct code comment inconsistencies.
% Compared to C4RLLaMA, which applies LLMs directly to code and comments, \tool provides a more structured and explainable solution.

Overall, \tool's approach is distinct in two aspects.
First, \tool specifically focuses on detecting inconsistencies in multi-parameter constraints, which is a missing piece in state-of-the-art works.
Next, \tool deals with code documentation, which involves longer and more complex text, and is more diverse than most code comments.

% to deal with this, \tool ultilize SOTA LLM and incorporates constraint fuzzy logic, which mitigates the uncertainty introduced by large language models.
% Last, \tool apply symbolic execution instead of AST-based program analysis, because~\xie{to check and fill; code comment vs documentation different;we specific on multiparameter; technical side, we .}

%% file: content/08.conclusion.tex
\section{Conclusion}\label{sec:conclusion}

In this paper, we propose \tool, a multi-parameter constraint checker for Python data science libraries. \tool utilizes both LLMs and symbolic execution to detect inconsistencies between code and documentation. 
To mitigate the uncertainty introduced by LLM outputs, we utilize constraint fuzzy logic to accommodate nearly-correct parameter constraints. The experimental results show that \tool is effective in identifying multi-parameter API documentation errors. We further reported 14 detected inconsistencies, 11 of which were confirmed by the development team.
Our work intuitively explores the multi-parameter constraint inconsistencies between code and documentation, and may inspire more future studies in this field.

\section{Data Availability}
The source code and dataset are available at \url{https://anonymous.4open.science/r/MPDetector-E30D}.

% We evaluated our tool on real-world popular Python data science and machine learning libraries and reported  to developers, with 11 inconsistency issues are confirmed.
% Our work intuitively explores the multi-parameter constraint inconsistencies between code and documentation, and may inspire more future studies in this field.
% proposed a customized solution for fuzzy constraint satisfaction problem to expand our constraint detection coverage. We evaluated our tool on real-world Python libraries and reported our findings to developers, with 11 inconsistency issues are confirmed.
% Our work intuitively explores the multi-parameter constraint inconsistencies between code and
% documentation, and may inspire more future studies in this field.

% \yi{we need a data availability section.}